\documentclass[12pt]{article}
\usepackage{graphicx,floatflt,amssymb,rotate} 
\input epsf.tex
\def\DESepsf(#1 width #2){\epsfxsize=#2 \epsfbox{#1}}

\newcommand{\be}{\begin{equation}}
\newcommand{\ee}{\end{equation}}
\newcommand{\br}{\begin{eqnarray}}
\newcommand{\er}{\end{eqnarray}}

\def\beq{\begin{equation}}
\def\eeq{\end{equation}}
\def\bnq{\begin{eqnarray}}
\def\enq{\end{eqnarray}}
\def\barr{\begin{array}}
\def\earr{\end{array}}

\def\be{\begin{equation}}
\def\ee{\end{equation}}
\def\ba{\begin{eqnarray}}
\def\ea{\end{eqnarray}}

\def\br{\begin{array}}
\def\er{\end{array}}
\def\bc{\begin{center}}
\def\ec{\end{center}}
\def\gsim{\mathop{\smash{>}}\limits_\sim}

\def\lapp{\mathrel{\rlap{\raise.5ex\hbox{$<$}}
                    {\lower.5ex\hbox{$\sim$}}}}
\def\gapp{\mathrel{\rlap{\raise.5ex\hbox{$>$}}
                    {\lower.5ex\hbox{$\sim$}}}}
\def\DESepsf(#1 width #2){\epsfxsize=#2 \epsfbox{#1}}
\usepackage{color}
\begin{document}
\thispagestyle{empty}
\vskip 30pt 
\begin{center}
{\bf Flavor unification, dark matter, proton decay and other observable predictions with 
low-scale $S_4$ symmetry}
\vskip .25in
{  Mina K. Parida${}^a$, Pradip K. Sahu${}^b$, Kalpana
  Bora${}^c$\\}
\vskip .15in
{\sl ${}^a$Harish-Chandra Research Institute,
Chhatnag Road, Jhusi, Allahabad 211019, India.}\\
{\sl ${}^b$ Institute of Physics, Sachivalaya Marg, Bhubaneswar 751005,
  India.}\\
{\sl ${}^c$ Department of Physics, Gauhati University, Guwahati 781014,
  India.}\\
\end{center}

\begin{abstract}
We show how gauge coupling unification is successfully implemented through  
non-
supersymmetric grand unified theory,
$SO(10)\times G_f (~G_f=S_4~, SO(3)_f,~ SU(3)_f)$,  
 using  low-scale  flavor symmetric model of the type
$SU(2)_L\times  U(1)_Y$ $ \times SU(3)_C \times S_4$ recently 
proposed by Hagedorn, Lindner, and Mohapatra, while assigning matter-parity
discrete symmetry for the dark matter stability. For gauge coupling
unification in the single-step breaking case, we show that a color-octet
fermion and a hyperchargeless weak-triplet fermionic dark matter are the 
 missing particles needed to complete its MSSM-equivalent degrees of freedom. 
When these are included the model automatically predicts the
nonsupersymmetric grand unification 
with a scale
identical to the minimal supersymmetric standard model/grand unified theory scale.  
 We also find a two-step breaking model with 
Pati-Salam intermediate symmetry where the dark matter and a low-mass color-octet
scalar or the fermion are signaled by grand unification.
 The proton-lifetime
predictions
are found to be accessible to ongoing or planned searches in a number of
models. We discuss grand unified origin of the light fermionic triplet dark matter, the
color-octet fermion, and their phenomenology.
\end{abstract}
\vskip 0.5 in
\date{\today}
\newpage

\section{Introduction}

The standard model (SM) gauge theory based upon $SU(2)_L\times U(1)_Y\times SU(3)_C$ 
($\equiv G_{213}$)
has enjoyed tremendous success by virtue of its excellent agreement with numerous 
experimental data. Nevertheless the SM has several shortcomings some of which are 
circumvented when the model emerges from a grand unified theory
(GUT) \cite{ps,georgi,so10}.
Apart from providing solutions on certain fundamental issues \cite{raby}, GUTs 
predict gauge boson mediated proton decay via $d=6$ operators 
and, in particular, the decay mode $p\rightarrow e^+\pi^0$ has been 
the hall mark of grand unification. Predictions of gauge boson 
mediated proton decays in nonsupersymmetric (non-SUSY) GUTs are neat 
and robust compared to the corresponding predictions in supersymmetric 
GUTs which are affected by  complications due to Higgsino mediated proton 
decays  via $d=5$ operators \cite{dim5, xxx}. Predictions  
on the proton decay in GUTs with or without supersymmetry (SUSY) 
have called for dedicated experimental searches to testify the predicted 
phenomena \cite{superk,dusel}. But GUTs may not provide a satisfactory answer
to fermion masses and mixings which may need 
additional  flavor symmetries.
In fact, experimental evidences of masses and large mixings of neutrinos  
has triggered interests in flavor symmetries leading to the suggestions of 
grand unification symmetry of flavor including $SO(10)\times G_f (G_f=S_4,~ SO(3)_f,~ SU(3)_f)$
 \cite{flav,hagedorn,leernm,parida,haibo,flav1,flav2,flav3}.

Recently Hagedorn, Lindner, and Mohapatra (HLM) \cite{hagedorn} have examined an interesting 
model based upon the non-SUSY  SM gauge structure as a possible solution to the 
fermion flavor problem with  $G_{213}\times S_4$ symmetry at low
scales. 
It predicts
a rich structure of neutral and charged Higgs scalars near the TeV or lower
scales which can be tested at Tevatron, LHC or planned accelerators. However,  suppressed
flavor-changing neutral current (FCNC) effects near electroweak scale may
suggest that the scale of spontaneous $S_4$ symmetry breaking could be higher
 ,$\sim$(few-10) TeV instead of being near the electroweak scale. In such a case
HLM type of analysis can be carried out with  renormalization group (RG)-extrapolated values of fermion masses and
mixings \cite{daspar}  at $1-10$ TeV scale as has been done in 
 \cite{leernm,parida,haibo} and in a
number of other models. Such HLM type of model with spontaneous $S_4$
braking at $\sim 1-10$ TeV would lead to the SM with only one
Higgs doublet below the TeV scale. 

In  SUSY $SO(10)\times S_4$  with  R-parity conserving  intermediate
symmetry  $SU(2)_L\times SU(2)_R\times U(1)_{B-L}
\times SU(3)_c\times S_4 (g_{2L}=g_{2R})$ at the type-I see-saw scale 
\cite{parida},
apparently there is no signature of the underlying flavor symmetry  to be
tested by accelerator searches.
While the  gauge hierarchy problem certainly
prefers a supersymmetric $SO(10)\times G_f$ model,
in the absence of any evidence of supersymmetry at low energies
 and for the sake of simplicity alone, 
prospects of minimal non-SUSY $SO(10)\times G_f$ should be thoroughly explored
and confronted with experiments.

More recently, interesting attempts have been made through non-SUSY $SO(10)$
 to exploit matter-parity origin  of dark matter (DM)
\cite{kadastik}.~As  signals of  
of grand unification in the single-step
breaking of non-SUSY $SO(10)$,  an inert scalar doublet along with a scalar  singlet
\cite{kadastik} have been suggested as DM candidates. In another independent 
study, a non-standard weak triplet fermion  
 $F_{\sigma}(3,0,1)$ with zero hyper-charge and TeV scale mass,
suggested earlier from phenomenological
 grounds \cite{cirelli}, has been identified as DM candidate \cite{frigerio}.   

In this work while attempting grand-unification completion of the HLM type
 model  we  identify two interesting
models  :(i) A single-step breaking model where $\sim$TeV scale masses of a fermionic triplet dark matter as well
as a color-octet fermion are predicted by grand unification; (ii) A two-step breaking model
with Pati-Salam intermediate symmetry where  TeV scale masses of the
 fermionic triplet DM and a color-octet
scalar or fermion are accommodated by  grand 
unification. We  show how light masses of both types of fermions
are obtained from the adjoint fermion representation 
$({45}_F,1)\subset SO(10)\times G_f$ and discuss their phenomenology. 
The proton lifetime predictions made in a number of the cases are accessible to ongoing and planned searches.
Although the 
 production cross section for direct detection of DM is known to be ~small at
 present LHC energies and  luminosity,
 there is agreement of recently predicted fluxes with PAMELA positron excess
with corresponding absence of anti-proton excess for energy $\le 200$ GeV.
  Large pair production cross
 section and absence of superpartners at accelerator energies would indicate
towards the presence of color-octet fermions of this model.
 Whereas  $SO(10)$ grand unification without flavor symmetry signals the presence of the fermionic
 triplet DM with TeV scale mass, the color-octet fermion needed for completion
 of the same grand unification has very large mass ($7\times 10^{10}$ GeV)
\cite{frigerio}
which is impossible to manifest at accelerator energies. In the present model,
however, both the fermion masses being in the $\sim$ TeV scale, are
 subject to experimental tests at the accelerators.\\

The non-SUSY $G_{213}\times S_4$ model with six doublets at lower scale has
an interesting prediction. Matching the degrees of freedom relevant for gauge
coupling unification with the minimal supersymmetric standard model (MSSM), we show that the color-octet fermion and the 
hyperchargeless weak-triplet fermionic DM are the missing non-trivial elements from the
low-scale flavor symmetric gauge theory. As such their inclusion at $\sim$TeV
scale  naturally predicts non-SUSY grand unification with a scale identical to
the MSSM-GUT scale.\\

This paper is organized in the following manner. In Sec. 2 we discuss 
briefly the HLM type model with $G_{213}\times S_4$ symmetry . 
In Sec.3 after showing absence of unification in the HLM type model,
the minimally modified single-step breaking scenario is presented with grand
unification signals. In this section we also discuss predictions 
on the proton lifetime. 
 In Sec.4 we discuss phenomenology of light fermions. 
The two-step breaking 
models including  Pati-Salam intermediate gauge symmetry 
 are discussed in Sec. 5.
Summary and conclusions are stated in Sec. 6.    

\section{ The standard gauge theory with low-scale $S_4$ symmetry}

In this section we discuss salient features of the $G_{213}\times S_4$ model of
the type used in ref. \cite{hagedorn} and briefly outline the HLM type of
model we have used to study possible signals of grand unification.

 In the non-Abelian discrete symmetry group $S_4$, there are two types of triplet 
representations, ${\bf 3_1}$ and  ${\bf 3_2}$, and also two types of singlet 
representations, ${\bf 1_1}$  and  ${\bf 1_2}$, but there is only one type of 
doublet representation, ${\bf 2}$. If one wishes to identify the fermions further in the fundamental 
representations of continuous flavor groups like  ${\bf SO(3)_f}$ or  ${\bf SU(3)_f}$,
then the three generations of standard fermions are to transform as ${\bf 3_2}$,
rather than ${\bf 3_1}$ of ${\bf S_4}$. The HLM \cite{hagedorn} proposal gives a very interesting 
possibility of embedding in the most attractive grand unified theory like $SO(10)$ 
by appending it with the continuous flavor group ${\bf G_f = SO(3)_f, 
 SU(3)_f}$ in addition to $S_4$. By fixing the Yukawa couplings to be symmetric in flavor space,
the allowed ${\bf SO(10)}$ Higgs representations are ${\bf 10}_H$'s  and 
${\bf {\overline {126}}_H}$'s. The sixplet of ${\bf S_4}$ doublet Higgs representations 
are appropriately fitted into those of ${\bf G_f}$ by using six ${10}_H$-plets of 
${\bf {SO(10)}}$ with transformation property 
${\bf ({10}_H, 3_1+2+1_1)}$ under ${\bf SO(10)\times S_4}$. In the minimal choice 
to create large right-handed Majorana neutrino mass term to drive the type-I see-saw 
mechanism, one additional Higgs representation transforming as  
${\bf (\overline{126}_H, 1)}$ under this group is needed \cite{type-I}. 
However in order to generate different masses of down quarks and charged 
leptons, five more ${\bf \overline {126}_H}$'s may be needed.
  Near the $S_4$ breaking scale, the model  consists of three generations of
fermions all transforming as $\bf{3_2}$ and six SM-like Higgs
fields transforming as $\bf{{1_1} + {2} + {3_1}}$
under $\bf{S_4}$. The particle content is summarized in Table \ref{tab1}. 

\begin{table}
\begin{center}
\begin{tabular}{|c|c|c|}\hline
Particle & ${\bf SU(2)_L\times U(1)_Y\times SU(3)_c}$  &
${\bf{S_4}}$ \\
\hline
Quarks $Q$ & $(\bf{2}, \bf{+{1\over 3}},\bf{3} )$ & $\bf{3_2}$\\
Anti quarks $u^c$ &  $(\bf{1}, \bf {-{4\over 3}},{\bf {\overline 3}} )$ & $\bf{3_2}$\\
Anti quarks  $ d^c$ &  $( \bf{1},+\frac{2}{3}, \overline{\bf{3}})$ & $\bf{3_2}$\\
Leptons $L$ & $( \bf{2}, -1, \bf{1} )$ & $\bf{3_2}$\\
Antileptons  $e^c$ & $(\bf{1},+2,\bf{1} )$ & $\bf{3_2}$\\
Right-handed $\nu$'s   & $(\bf{1},0,\bf{1} )$ & $\bf{3_2}$\\
\hline
Doublet Higgs  $\bf \phi_0$ & $( \bf{2}, -1,\bf{1} )$ & $\bf{1_1}$\\
Doublet Higgs $\bf (\phi_1,\phi_2)$ & $(\bf{2}, -1,\bf{1}  )$ & $\bf{2}$\\
Doublet Higgs $\bf (\xi_1,\xi_2,\xi_{3})$ & $( \bf{2}, -1,\bf{1}  )$ & $\bf{3_1}$\\
\hline
\end{tabular}
\end{center}
\begin{center}
\begin{minipage}[t]{12cm}
\caption[]{ The minimal particle content and transformation  properties
in the ${\bf G_{213}\times S_4}$ model used in Ref. \cite{hagedorn}.
\label{tab1}}
\end{minipage}
\end{center}
\end{table}
\vskip 0.1 in
Fermion masses and mixings in the model have been derived through a 
 ${\bf G_{213} \times {S_4}}$ invariant Yukawa
Lagrangian explicitly given in ref. \cite{hagedorn}.
Although there are twelve Yukawa couplings 
in the $G_{213}\times S_4$ model, when embedded in a GUT like ${\bf SO(10) \times S_4}$ 
they are expected to reduce to only three 
corresponding to three of its representations ${\bf ({10}_H,~ 1_1)}$  ${\bf ({10}_H,~ 2)}$ 
and ${\bf ({10}_H,~ 3_1)}$. These three are expected to reduce to only one 
if the discrete flavor symmetry group emerges from continuous flavor group
$G_f = SO(3)_F, SU(3)_f$.
Although some of the CKM-quark mixings have been found to be somewhat smaller than the 
experimental values, in two 
numerical examples, the charged fermion masses have been  shown to arise as small 
deviations from well known rank one matrices.

Although the model has the potentiality to accommodate  type-II seesaw 
\cite{vseesaw}, the $SU(2)_L$  Higgs triplets contained in
$(\overline{126}_H,~ 1)$ are excluded to reduce the number of parameters.  
The right-handed (RH) Majorana neutrino mass matrix generated by $(\overline{126}_H,~ 1)$ is proportional to 
$3\times 3$ unit matrix in the 
generation space which drives the type-I see-saw formula for light neutrino 
masses \cite{type-I, vseesaw} and the HLM model requires the quasi-degenerate RH
neutrino mass scale to be $M_R\simeq 10^{13}$ GeV. Thus the model has high
potentiality to explain baryon asymmetry of the universe through resonant
leptogenesis and in non-SUSY models  there is no gravitino constraint \cite{mkpar}. 

Although spontaneous $S_4$ breaking scale in the HLM model has been assumed to
 be near the electroweak scale, suppression of flavor-changing
 neutral current (FCNC) may require all nonstandard Higgs
 doublet masses to be $\sim$ O(TeV) or larger. In that case using RG-extrapolated values
 of fermion masses and mixings \cite{daspar} HLM-type of
analysis can be carried out in the $G_{213}\times S_4$ model  with  spontaneous $S_4$ breaking at $\mu \sim O(1-10)$ TeV
 leading to the SM  with only one light Higgs doublet at lower
 scales. Fits to the extrapolated values of masses and mixings 
have been carried out at $\mu=10^{13}-10^{15}$ GeV in 
                  \cite{leernm,parida,haibo,altbank} and in a number of other models.
Also since the extrapolation is to be done to a scale which is only $\sim 1-2$ 
order larger than the electro-weak scale, the numerical results are expected 
to be similar to the HLM fit with small differences. In any case for
 studying grand unification and capturing  new signals for low-scale
 physics, no numerical inputs from fits to fermion masses and mixings are
 needed either from the HLM model \cite{hagedorn} or from its possible type
with $(1-10)$ TeV $S_4$ symmetry breaking scale.  

 In what follows for the purpose of embedding the HLM type model in $SO(10)\times G_f$ we will assume that 
  $G_{213}\times S_4$ symmetry breaks to SM softly or spontaneously
 at $\mu= M_S \sim (1-10)$ TeV leading to the SM with only the standard Higgs
 doublet below the TeV scale.

\section{ Unification through single-step breaking and matter parity
  conservation}

While implementing coupling unification, another purpose of the present 
work is to identify particles with 
$\sim$ TeV scale masses  as signals of
grand unification which may be cold dark matter (CDM) candidates of
the universe \cite{cirelli,kadastik,frigerio} or other nonstandard particles
accessible to collider searches. The origin of discrete
symmetry  existing in  $SM\times S_4$ model or in the SM itself which ensures 
DM stability is discussed below.
  
\subsection{Matter parity conservation in $SO(10)\times G_f$ breaking}

It has been known for quite some time that matter parity is a
discrete symmetry of the standard model  \cite{rnmpar},
\ba
 P_M= (-1)^{3(B-L)},\nonumber \label{matterp}
\ea
 where $B(L)$ is the baryon(lepton) number. $(B-L)$ is an
element of gauge transformation in $SO(10)$
and it is  the $15$th generator
 of Pati-Salam color-gauge group  $SU(4)_C$. Also when 
$SO(10)\to SU(5)\times U(1)_{\chi}\to 
SM$,   $\chi = 4T_{3R}+3(B-L)$ and since $4T_{3R}$ is always even, the
 $\chi$-parity , $P_{\chi}=(-1)^{\chi}=P_M$.  
 Matter parity survives as a discrete symmetry provided 
the symmetry breaking of $SO(10)$ to the SM undergoes through Higgs scalars
carrying even $(B-L)$ which  explains tiny left-handed (LH) neutrino masses through type-I
seesaw mechanism.
The survival of matter parity as a discrete symmetry in the SM also follows
 from general arguments for even $B-L$ \cite{wilczek,martin}.    
The vacuum expectation value (VEV) of the right-handed (RH) Higgs triplet
carrying (B-L)= -2 is  contained in 
$({\overline{126}_H}, 1)$ of $SO(10)\times G_f$. This has been utilized in  
 one-step or
two-step breaking models discussed throughout this work to ensure survival of
matter parity. 
In $SO(10)$, for dimension of representations $\le 210$, while the representation $16$ and 
$144$ have odd $P_M$, the matter parity of representations $10, 45, 54, 120,
 126, {\overline {126}}$ , and $210$ is even.
Utilizing this property  and, as necessary requirement of
 gauge coupling unification, non-standard Higgs scalars
(singlet and neutral component of weak doublet) in ${16}_H$ \cite{kadastik} or neutral
 component of non-standard weak triplet fermion contained 
in ${45}_F$ \cite{frigerio} have been identified as possible DM candidates.
  While a fermionic color-octet has been also found necessary for coupling
  unification in \cite{frigerio}, it can never be directly observed at accelerator energies
  because the predicted value of its mass is large ($7\times 10^{10}$ GeV). In 
contrast, while searching for completion of grand
  unification in the presence of flavor symmetry, it will be shown in this work
that both the triplet DM and the color-octet fermion 
with $\sim$ TeV scale masses are signaled by grand unification and, as such,
both are accessible to the ongoing or planned accelerator searches.
  
\subsection{Absence of unification in the minimal model}

\begin{figure}[htb]
\begin{center}
\includegraphics[width=12cm]{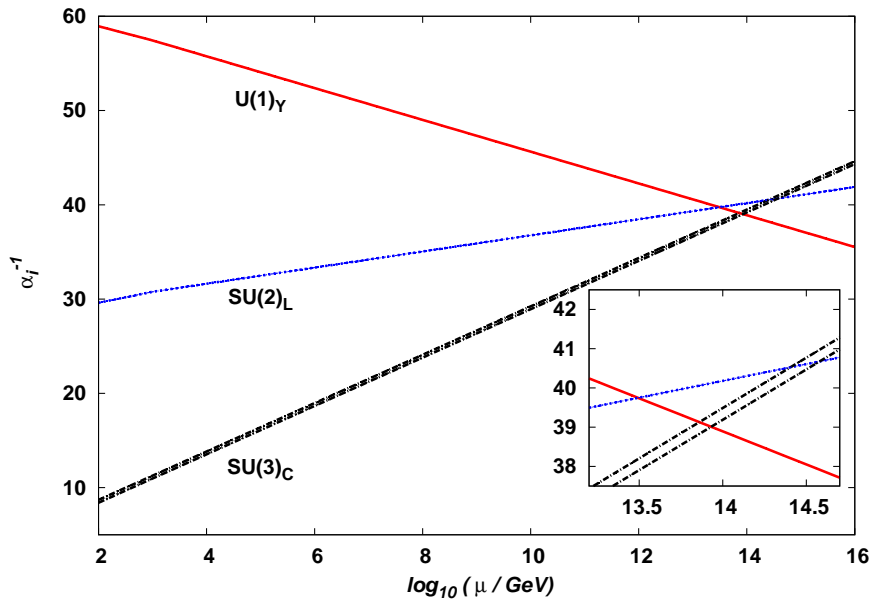}
\caption{RG evolutions of three gauge couplings of the standard model from $\mu= M_Z$
to $\mu = M_{\rm{Planck}}$ with $M_S\simeq 1$ TeV 
where $\alpha_S(M_Z)$ has been used with $3\sigma$ uncertainty and others with $1\sigma$
uncertainty.  The inset in the figure shows the presence
 of the triangular region at the high scale where the two dot-dashed parallel
 lines represent the $3\sigma$ boundaries of  $\alpha^{-1}_S(M_Z)$.
}\label{Fig1}
\end{center}
\end{figure}

 In this section  we search for gauge coupling unification of the HLM
type model with six electroweak doublets belonging to the ${\bf S_4}$ 
representations ${\bf (3_1+2+1_1)}$.
We assume  the symmetry 
${\bf G_{213}\times S_4}$ to be restored at  $\mu = M_S\sim (1-10)$  TeV 
and, thereafter, to continue till $SO(10) \times G_f$ symmetry 
takes over at the GUT scale 
$\mu = M_U \gsim 10^{15}$ GeV with no intermediate gauge symmetry.
Between the scales 
$M_Z$ and $M_{S}$,
the standard model symmetry with one Higgs doublet is assumed to operate. For this
Model I we consider 
\ba
{\bf SO(10) \times G_f} ~~\frac{M_U}{}~~~{\bf G_{213}\times 
S_4} ~~\frac{M_{S}{}}~~~{\bf G_{213}}. \label{1step}
\ea
where we give GUT-scale vacuum expectation values (VEVs) to the relevant components of Higgs representations $(54_H,1)\oplus (45_H,1) \oplus 
({\overline{126}_H},1)$ under $SO(10)\times G_f$ in the first step of symmetry
breaking. The second and subsequent steps proceed in a similar manner as explained in \cite{hagedorn}. 
For the evolution of gauge couplings we utilize the two-loop renormalization 
group equations \cite{gqw,jones},

\ba
{\mbox{d} \alpha_{i} \over \mbox{d} t} = {\mbox {a}_{i} \over 2\pi} \alpha_i^{2} + 
\sum_{j}{\mbox{b}_{ij} \over 8 \pi^2} \alpha_{i}^{2} \alpha_{j}.
\ea
In our notation $a_i~ (i=1, 2, 3)$ denote one-loop beta function coefficients for the fine-structure 
constants 
of $U(1)_Y,~ SU(2)_L$ and $SU(3)_C$ , respectively, and $b_{ij}, (i, j = 1, 2, 3)$ denote 
the corresponding two-loop 
coefficients as elements of a $3\times 3 $ matrix. 
Noting that $a_i= (41/10, -19/6,$ $-7)$ for the SM with one doublet, but
$a_i = (23/5, -7/3, -7)$ in the $G_{213}\times S_4$ model with six doublets,
we have used the Particle-Data-Group values\cite{PDG10} of  
 $\sin^2\theta_W(M_Z)=0.23116\pm 0.00013$ and $\alpha^{-1}(M_Z)=127.9$. In
 order to make the point
more convincing on whether unification is  taking place in the minimal model, we have chosen 
$3\sigma$
deviation  from the global average of the strong interaction coupling, $\alpha_S(M_Z)=0.1184
 \pm 0.0007$, so that, statistically, our result would be valid at   
 $99.7\%$ confidence level.
The evolutions of the three 
gauge coupling-constants from $\mu= M_Z$
to $\mu = M_{\rm {Planck}}$ is shown in Fig. \ref{Fig1} where the widths of
electroweak lines are at $1\sigma$ but the width of strong-interaction
 coupling is at
$3\sigma$. 

When the $S_4$ spontaneous breaking scale is changed from $M_S= 1$TeV to $10$
TeV there is no significant change of the triangular region. 
In particular when this scale becomes large with $M_S\simeq 10^{14}$ GeV, the
triangular structure of the minimal non-SUSY SM appears exhibiting the
well known deconstructed unification.
Although the size of the triangle appears to be smaller when the $3\sigma$ 
error bar in  $\alpha_S(M_Z)$ is taken into account, the non-overlapping region is prominent to show 
 that  
the inverse fine structure constants cross at three 
different points. Evidently there is no possibility of gauge coupling unification with the minimal particle 
content of the $G_{213}\times S_4$ model.

\subsection{ Unification with fermionic triplet dark matter and color-octet fermion}

It has been shown in a number of investigations in the absence of flavor
symmetry in supersymmetric as well as non-supersymmetric models, with or
without intermediate symmetry, that grand unification of gauge 
couplings at 
$M_U\geq 10^{15}$ GeV is achieved
provided there are additional particle degrees of freedom (scalars or fermions 
) at lower scales \cite{kadastik,frigerio,fram,manohar,
  dorsner,lm,pnath,perez,mkpar}. 
The light scalars  needed for completion of grand unification
require additional fine tuning and 
the criteria of minimal fine tuning \cite{min} are to be relaxed.
As we are discussing  unification along with flavor group through $SO(10)\times G_{f}$, there is
also the possibility that a single additional fine tuning that would have made 
one submultiplet of a GUT-representation light without flavor symmetry, would
now make a n-tuple of $G_f$ light. 
 On the other hand if unification is achieved with non-standard light
 fermions, there may be a global $U(1)$ symmetry to protect their masses and
 the fine-tuning may not be so unnatural.
 Although
finally we achieve unification with light fermions only, we start with light scalars.

Using one-loop and two-loop coefficients in different mass ranges 
we  find  unification is possible with  
a pair of color-octet scalars $2C(1, 0, 8)$, and a pair of weak triplet
scalars $2\sigma(3, 0, 1)$, with both the masses near the TeV scale and 
a GUT scale suitable to guarantee observed proton stability 
\footnote{We have
  checked that solutions to coupling unification obtained in all cases discussed in this paper can
  also be obtained if the $S_4$ symmetry breaking scale is larger, 1-10
  TeV, needed to avoid more than one
 Higgs doublets at lower scales, suppress FCNC effects and Higgs search
  prospects at Tevatron and  LHC.}

\ba
& & M_{S}=10^{2.5}~~ GeV,~~ M_{X}=M_{\sigma}(3,0,1)=10^{3}~~ GeV, \nonumber \\
& & M_C(1,0,8)=10^{3.5}~~ GeV,~~ M_{U}=10^{16}~~ GeV,~~ 1/\alpha_G=35.3.
\ea

\begin{figure}[hbt]
\begin{center}
\includegraphics[width=10cm]{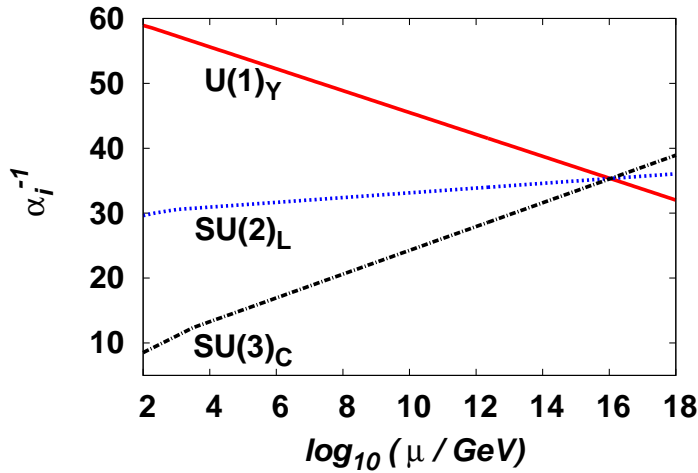}
\caption{ Unification of gauge couplings in single-step breaking 
Model-I with scalars $2\sigma(3, 0, 1)$ and $2C(1, 0, 8)$ at $M_{\sigma}
\simeq 1$ TeV and $M_C\simeq 3$ TeV, respectively. The unification pattern is
unchanged when all the four scalars are replaced by fermions
$F_{\sigma}(3,0,1)$ and  $F_{C}(1,0,8)$ at the respective scales.
}\label{Fig2}
\end{center}
\end{figure}

For this model almost exact unification of the three gauge couplings is shown 
Fig.\ref{Fig2}.

In this Model I, the pairs of Higgs scalars $2C(1, 0, 8)$ and $2\sigma(3, 0, 1)$
can be embedded into $({210}_H, 2)$, or $({45}_H, 2)$ under $SO(10)\times S_4$, or
into the representation $({24}_H, 2)$ under $SU(5)\times S_4$. 

Purely from the requirement of coupling unification, this leads to  
 an interesting possibility of replacing each pair by the corresponding
 fermions, $F_{\sigma}(3,0,1)$ and  $F_{C}(1,0,8)$ at the same scale.
The simple reason for this possible replacement is that the one-loop beta
function coefficient remains  the same as the scalar case leading to
almost the same pattern of unification as in Fig.\ref{Fig2}. The
fermionic weak-triplet at the TeV scale which is a color singlet 
and has hypercharge Y=0  can be identified as  a 
dark matter candidate if it does not have Yukawa interaction with standard
model particles.

These results arrived through numerical analyzes have a simple
analytic derivation.
Analyzed in a straight-forward manner, the above results turn out to be
automatic 
predictions of the  $G_{213}\times S_4$ model  with six doublets. The proof
proceeds through the following steps.
The first observation is that, with respect to one-loop contributions to gauge
couplings, the six doublets are equivalent to two Higgs
doublets and their superpartners of the MSSM. The second observation is the
well known fact that the scalar superpartners of quarks and leptons do not 
determine the MSSM GUT scale although they contribute to the value of the GUT
gauge coupling.~Using these two observations, it immediately follows that 
there are only the following nontrivial  degrees of freedom missing from the
low-mass non-SUSY spectrum to match the MSSM spectrum sans squarks and sleptons: the octet fermion and the triplet fermionic
DM. When these degrees of freedom are included in the non-SUSY model near the
TeV scale, it predicts
grand unification of gauge couplings with the non-SUSY GUT scale identical to the
MSSM-GUT scale($\simeq 10^{16}$ GeV)\cite{mssmgut}. Both the GUT scale and the
predicted low-mass particles are specific to this non-SUSY six-Higgs doublet
model which have been also obtained by independent numerical analyzes stated above.

Noting that the quantum numbers of the weak-triplet and the 
color octet match the corresponding components of the 
adjoint representation, we suggest that these fermions are lighter components 
of the non-standard fermionic representations  $(45_F, 1)\subset SO(10)\times
S_4$. As this representation has even matter parity, it does not couple to 
 standard model fermions or the Higgs scalar directly, although there could be
 matter-parity conserving non-standard  Yukawa interaction which has
 been discussed
 in Sec.4 in some model extensions. Even in the $SO(10)$ theory itself
${45}_F$ does not have usual Yukawa interaction with standard fermions in
${16}_F$ through SM Higgs doublets which might originate from ${10}_H, 
{\overline {126}_H}$, and ${120}_H$ or their linear combinations.   

It is worthwhile to compare fermionic signals of grand unification between the
conventional $SO(10)$ \cite{frigerio} and the present model. In the $SO(10)$
model in addition to the triplet fermionic dark matter near the TeV scale,
a color-octet fermion with  mass $7\times 10^{10}$ GeV was also
needed to complete grand unification. Because of the high mass it appears
impossible to testify the presence of such a fermion at accelerator energies.
In the present model , apart from components of additional  Higgs doublets with masses  $\sim$ few TeV
- 10 TeV,  which are natural ingredients of the model, completion of grand
unification predicts both the triplet fermionic dark matter as well as the
color-octet fermion with $\sim$ TeV scale masses; as such they are subject to
verification at accelerator energies. In Sec.4 we will
discuss some phenomenological consequences of these light fermions while we 
derive their
TeV scale masses using renormalizable Yukawa Lagrangian of  $(45_F, 1)\subset SO(10)\times G_f$.    

Neutrino masses and mixings being governed by the type-I see-saw mechanism,
the RH neutrino mass matrix is proportional to a diagonal matrix. As  
 ($\overline{126}_H, 1$) generates the RH neutrino mass via its coupling
$f~16_f~16_f~{\overline {126}_H}$, and the vacuum expectation value 
$<{\overline{126}_H}>
\sim M_{GUT}$, it is necessary that the Majorana type Yukawa coupling 
 $f\simeq 10^{-3}$ in Model I.
In the next subsection we estimate proton lifetime predictions in this  model
by including uncertainties due to GUT-threshold as well as low-scale threshold
effects \cite{thrs1,thrs2,thrs3,lmpr} in order to have an approximate idea of the allowed
range for experimental accessibility.

\subsection{ Predictions on proton lifetime }

In $SU(5)$ model the $d=6$ proton decay operator emerges from 12 superheavy
gauge bosons contained in $(2,-\frac{5}{3}, 3)\oplus 
(2,\frac{5}{3},\overline{3})$ under $G_{213}$. In $SO(10)$ the superheavy 
gauge bosons transform simultaneously as LH and RH doublets 
and are contained  
in the $(2,~ 2,~ 6)$ multiplet under Pati-Salam group $G_{224}$.  Up to
a good approximation, the decay width for $p\rightarrow e^+\pi^0$ in all
models investigated in this work can be written as \cite{pnath,bajc,mkpar}

\ba 
\Gamma(p\rightarrow e^+\pi^0) \nonumber
&=&\frac{m_p}{64\pi f_{\pi}^2}
(\frac{{g_G}^4}{{M_U}^4}){A_L}^2\bar{\alpha_H}^2(1+D+F)^2[(A_{SR}^2+A_{SL}^2)\\
&\times & (1+ |{V_{ud}}|^2)^2]
\label{eq9}
\ea

In eq.(\ref{eq9}) $M_U$ represents degenerate mass of $24$ superheavy gauge
bosons and $g_G$ is their coupling to quarks and leptons 
($\alpha_G=g_G^2/{4\pi}$) at the GUT scale $\mu=M_U$. 
Here $\bar\alpha_H$= hadronic matrix elements, $m_p$=proton mass=938.3 MeV, $f_{\pi}$=pion decay 
constant =139 MeV, and the chiral Lagrangian parameters are D=0.81 and F=0.47. $V_{ud}$ 
represents the CKM- matrix element $(V_{CKM})_{12}$ for quark mixings.

The dimension $6$
operator when evolved down to the GeV scale, short-distance renormalization factor from 
$\mu=M_U - M_Z$ turns out to be 
$A_{SL}\simeq A_{SR}\simeq A_{SD} \simeq 2.566$  for Model I and the long distance 
renormalization factor is $A_L\simeq 1.25$.
These are estimated using values of gauge couplings in the relevant mass ranges, the anomalous
dimensions  and the one-loop beta-function coefficients \cite{buras}. Using 
$A_R=A_L A_{SD}\simeq 3.20$,
$F_q=2(1+|V_{ud}|^2)^2\simeq 7.6$, we express inverse decay width for $p\rightarrow e^+\pi^0$ as

\ba
\Gamma^{-1}(p\rightarrow e^+\pi^0)\nonumber
& = &
1.01\times10^{34} yrs. \left[\frac{0.012~{\rm GeV}^3}{\alpha_H}\right]^2\left [\frac{3.2}{A_R}\right ]^2
\\ 
&\times &\left [\frac{1/35.3}{\alpha_G}\right ]^2\left [\frac{7.6}{F_q}\right ]\left 
[\frac{M_U}{4.659\times 10^{15}}\right ]^4.
\label{gama} 
\ea

where we have used $\alpha_H ={\bar {\alpha_H}}(1+D+F) \simeq 0.012$ GeV$^3$
from recent lattice theory estimations\cite{lattice}. 
\vskip 0.1 in
Using the two-loop estimations of  Model I with  $M_U=M_U^0=10^{16}$ GeV in 
eq.(\ref{gama}) gives\\ 
\be
\tau_p^0=2.48\times10^{35}~~ yrs.
\ee
which is nearly 25 times longer than the current 
experimental limit, but
accessible to measurements by next generation proton decay searches.\\

\subsection{Threshold effects}

We have found that completion of grand unification requires a pair of
weak-triplet scalars and a pair of color-octet scalars with masses near the
TeV scale which could be members
of $S_4$-doublets. Alternatively, the same unification is completed by
fermionic weak-triplet as a prospective DM candidate and a fermionic
color-octet, both  with $\sim$TeV scale masses.
The origin of these additional low-mass
scalars may be attributed to the adjoint representations $({45}_H,2)$ , or 
equivalently,
the low-mass fermions may originate from $({45}_F,1)$
under $SO(10)\times G_f$. Although the representations
${54}_{F,H}$ or ${210}_{F,H}$ may be chosen instead of ${45}_{F,H}$,
we prefer to choose the the smallest representation among the three.
We find that contributions of superheavy components of  $({45}_{F,H})$
towards  GUT-threshold correction on unification  
mass vanishes and its possible reason has been explained \cite{rnmtheo} . Including the GUT-threshold effects of non-degenerate\footnote{This non-degeneracy is not the same as in the 
conventional sense. In this and all other  models, we have assumed that
all superheavy sub-multiplets belonging to any particular GUT representation are degenerate 
in mass whereas there could be non-degeneracy among masses assigned to different GUT 
representations.} 
superheavy
components in  ${\overline {126}_H}$ and
$6({10}_H)$, and low-scale $M_S-$ threshold effects  due to six light Higgs
doublets treated as degenerate,
the maximal uncertainties on the unification mass and proton lifetime are
\cite{thrs1, thrs2, thrs3},

\ba
\frac{M_U}{M_U^0}&=& 10^{\pm 0.0465|\eta|\pm 0.031|\eta^{\prime}|},\nonumber\\ 
\ea

where  $\eta =\ln(M_{SH}/M_U)$, $M_{SH}$ being the scale of superheavy masses
lighter or heavier than
 the GUT-scale. It characterizes splitting of masses around $M_U$. Here
 $\eta^{\prime}=\ln(M_D/M_S)$, $M_D$ being the common mass of degenerate
doublets around $M_S-$ threshold. We have noted that maximal contribution to 
uncertainty
due to $M_S-$ threshold occurs when the doublets are degenerate.    

Using eq.(6) and eq.(7), we find that even for non-degenerate  superheavy component masses
$10(\frac{1}{10})$ times heavier(lighter) than the GUT-scale corresponding to
$|\eta| = \ln (10)$  and $M_S-$ 
threshold parameter 
$|\eta^{\prime}|=\ln (5)$, the predicted proton lifetime is 

\be
\tau_p=10^{35.245\mp 0.428\pm 0.2}~~ yrs,
\ee
which is in the accessible range of planned searches \cite{dusel}.

\section{Masses of non-standard fermions and  phenomenology}

We have noted that a weak triplet fermion  $F_{\sigma}(3,0,1)$ and a
color-octet fermion  $F_C(1,0,8)$  with $\sim$ TeV scale masses are 
predicted by flavor-symmetric grand unification in single-step breaking  
Model I . 
The presence of such low mass fermions  may not
be as unnatural since, in the limiting case of their vanishing masses, they may be
protected by corresponding $U(1)$ global  symmetry.   
In this section, using  
 $SO(10)\times G_f$ theory, we show how they can be light  and  briefly discuss
their phenomenology .
We introduce the adjoint fermion representation $A_F=(45_F, 1)$ 
and the Higgs representations $E({54}_H,1)$ and $\Phi({210}_H,1)$ under
 $SO(10)\times G_f$ and consider the renormalizable Yukawa Lagrangian at the GUT scale,
\ba
  -L_{Yuk} = A_F\left(m _A + h_p\Phi +h_e E\right)A_F.\nonumber
\label{Yuk}
\ea
with $m_A\simeq M_U$.
While $E$ has only one  singlet, $\Phi$ has three singlets 
$\Phi_i(i=1,2,3)$ under SM. When GUT-scale VEVs are assigned to $E$ and $\Phi$ 
 along these directions ,besides the GUT
symmetry breaking, the fermion
components  in $A_F$ get masses \cite{fuku},

\ba
m(1,2/3,3)&=& m_A+{\sqrt 2}h_p{\Phi_2 \over 3} -
2h_e{<E>\over {\sqrt {15}}}, \nonumber \\
m(2,1/6,3)&=& m_A + h_p{\Phi_3\over 3} +h_e{<E>\over {2
  \sqrt{15}}},\nonumber \\
m(2,-5/6,3)&=& m_A - h_p{\Phi_3\over 3} +h_e{<E>\over {2
  \sqrt{15}}},\nonumber \\  
m(1,1,1)&=& m_A +{\sqrt {2}}h_p{\Phi_1\over {\sqrt {3}}} +
{\sqrt {3}}h_e{<E>\over  {\sqrt {5}}},\nonumber \\ 
m(1,0,1)&=& m_A +{\sqrt {2/3}} h_p\Phi_1 +
{\sqrt{3/5}}h_e<E>,\nonumber \\
m^{\prime}(1,0,1)&=& m_A +{2\sqrt {2}\over 3} h_p\Phi_2 -
{2\over \sqrt{15}}h_e<E>,\nonumber \\ 
m_{F_C}(1,0,8)&=& m_A -{{\sqrt {2}}\over 3} h_p\Phi_2 -
{2\over {\sqrt{15}}}h_e <E>,\nonumber \\ 
m_{F_{\sigma}}(3,0,1)&=& m_A -\sqrt{2\over 3} h_p\Phi_1 +\sqrt{3\over
  5}h_e<E>.\nonumber
\label{masses}
\ea

It is clear that by tuning any two of the parameters 
while the weak triplet and color-octet fermion masses, $m_{F_{\sigma}}(1,3,0)$ and $ m_{F_C}(1,0,8)$, are brought to the $\sim$TeV scale, all
other components have masses near the GUT scale. These two fermions are
analogous to wino and gluino of split-SUSY models where the scalar
superpartners have very large masses \cite{arkani}. Although from the
minimality of the dimension of representation, we have chosen $({45}_F, 1)$ as
the possible origin of the two light fermionic submultiplets, alternatively,
$({54}_F,1)$ or $({210}_F,1)$ may be chosen with similar derivation.
             
\par\noindent{\bf (i) Weak-triplet fermionic dark-matter}
                   
 The decay of the heavier charged components $F^{\pm}_{\sigma}$ to the lighter 
neutral component via weak-gauge interaction
   leads to the mass difference  ~$m_{F^{\pm}_{\sigma}}- m_{F^0_{\sigma}} = 166$ MeV \cite{cirelli,dmph0}. 
 Within $3\sigma$ uncertainty of the WMAP data on relic density its mass 
 has been estimated as  $m_{F_{\sigma}} =2.75 \pm 0.15$ TeV
 corresponding to the Sommerfeld resonance value at $2.5$
TeV \cite{cirelli,dmph1}. In a more recent analysis, taking into account the effect
of kinetic decoupling, the Sommerfeld resonant value has been found to be 
the same as the triplet mass  $m_{F_{\sigma}} \simeq 4.5$ TeV \cite{subhendra}. 

  Elastic scattering of DM off the nucleon occurs through the loop-mediated
 W-boson exchange with and without the SM Higgs boson and
leads to a suppressed spin-independent cross section \cite{cirelli}. Although this
cross section is $2-3$ orders of magnitude 
lower than the current experimental sensitivities, it is  
expected to be within the accessible range of planned
experiments for direct detection \cite{xenon}. The 
large mass splitting between the  charged and neutral
components of the triplet ($\simeq 166$ MeV) compared to  the proton-neutron
mass difference or the DM kinetic energy , kinematically  forbids inelastic scattering. 
 
For indirect detection,  DM pair
 annihilation  and resulting  fluxes of photons, antiprotons, and positrons,
 diffuse or from the center of Milky-Way galaxy, have been
 predicted \cite{cirelli, dmph1,dmph4,subhendra}.  Because of the proximity of the highly
 nonrelativistic triplet mass 
to the Sommerfeld resonant value, the DM-annihilations are boosted resulting
 in enhancement by a factor as large as 
$\sim O(100)$. 
 The recent estimation with  $m_{F_{\sigma}} \simeq 4.5$ TeV  has explained 
the observed PAMELA excess of positrons \cite{pamela} boosted by Sommerfeld 
resonance effect. The anti-proton flux prediction agrees well with
the present measurement even up to energies $\le 200$ GeV \cite{antiproton}; but  a clear trend of
this boosted flux is predicted in the region of $300-1000$ GeV in which no
experimental data are yet available \cite{subhendra}.      

Experiments using atmospheric Cherenkov telescopes expect to observe
 monochromatic photons with energy $\simeq m_{F_{\sigma}}$ which originate from
pair annihilation $F_{\sigma}{\bar F}_{\sigma}\to \gamma\gamma $ \cite{dmph1,dmph4}. Observation of such
photons would determine the triplet mass but non-observation would rule out
 the triplet-DM model. More recent analysis of first two years of  Fermi Gamma Ray
 Space Telescope data from galactic center have been found to fit a low mass 
DM particle in the range $7-10$ GeV. \cite{hooper}.

The $2.75$ TeV triplet DM production rate for direct detection at colliders
 has been estimated to be $\sim 10^{-45}$ cm$^2$ which is accessible with
 improved accelerator energy and  luminosity.
 With  LHC  luminosity of $100$
fb$^{-1}$, the $pp\to F^+_{\sigma}F^-_{\sigma}X$ cross section has been predicted to
produce only one DM pair \cite{cirelli}. On the other hand, if a hadronic collider is available with twice
the LHC energy, then it  will have production cross section and produced number of DM
pairs several orders of magnitude
larger. In $e^+e^-$ collider with energy $\simeq 5.5- 8$ TeV, observation of
$F_{\sigma}^+F_{\sigma}^-$ pair is possible   
 through Z-exchange at tree level while $F_{\sigma}^0{\bar{F}}_{\sigma}^0$ pair production is allowed 
at at loop level. 
After production it may be easier to identify the charged
component of the triplet as it is predicted to leave long lived tracks
corresponding to the estimated lifetime of $\simeq 5.5$ cm. The alternative
 possibility with a large mass ($4.5$ TeV) for the triplet appears to cause 
problem for direct production and detection at LHC energies.
However, if the phenomenon of cold dark matter originates from 
more than one components including the triplet, then the triplet mass may
be smaller and easier for collider signatures but at the cost of predictive
power of the model.
 
\par\noindent{\bf (ii) Color-octet fermion }

In Model I, in addition to the fermionic weak-triplet, completion of gauge
coupling unification has been found to require the presence of color-octet fermion $F_C(1,0,8)$
or, equivalently, a pair of color-octet scalars $C(1,0,8)$ and we have
suggested their possible origins from the adjoint  
representations $({45}_F, 1)$ or $({45}_H , 2)$ of $SO(10)\times G_f$ .
Here at first we discuss briefly the more interesting case of the color-octet fermion. Consistent gauge coupling
unification in  Model I  
 is noted to be possible for rather a wider mass range of the color-octet 
fermions $m_{F_C}(1,0,8)=500$
 GeV $- 10$ TeV.

Being hadron colliders, both Tevatron and LHC are expected to show much higher
rate of production of  color-octets compared to color-triplets because of
larger Dynkin index. The production of ${F_C}(1,0,8)$  at hadron collider would be in pairs via
gluon-gluon fusion or through $q{\bar q}^{\prime}$ annihilation in a manner
similar to gluino pair production. 
In the leading order(LO) using the parton level amplitudes 
   for $F_C{\bar {F}_C}$ production in the non-SUSY case via
  quark-antiquark annihilation or via  
gluon-gluon fusion   
 \cite{sekhar}, the parton level cross sections are,

\ba
{\hat {\sigma}}(q{\bar q}^{\prime}\to F_C{\bar {F}_C}) &=&
\frac{8\pi\alpha_S^2}{9{\hat s}}
[(1+\kappa/2)(1-\kappa)^{1/2}], \nonumber\\
{\hat {\sigma}}(gg\to F_C{\bar {F}_C}) &=& \frac{9\pi\alpha_S^2}{32{\hat s}}
[(8+4\kappa+2\kappa^2)\ln\frac{1+(1-\kappa)^{1/2}}{1-(1-\kappa)^{1/2}}
-\frac{2}{3}(16+17\kappa)(1-\kappa)^{1/2}].\\ 
\nonumber
\label{pcross}
\ea
where ${\hat s}=$ partonic c.m. energy squared and $\kappa = 4m_{F_C}^2/{\hat s} \le 1$.

Using CTEQ6 parton density distribution
function \cite{pumplin} and integrating , we obtain the total pair production cross
section $\sigma({\rm pb})$  at LHC energy of
$\sqrt s = 14$ TeV  and the
number(N) of  $F_C{\bar {F}_C}$ pairs produced for different values of
color-octet fermion masses,

\ba
\sigma ({\rm pb}) &=& 0.9,~~~~~~~~~~~~1.0\times 10^{-2}, ~1.5\times 10^{-4},~2.5\times
10^{-5},\nonumber\\ 
 N           &=&9\times 10^4,~~~1\times 10^3,~~~~~15,~~~~~~~~~~~2.5,
\nonumber\\ 
 m_{F_C}({\rm TeV})&=& 1.0,~~~~~~~~~~~2.0,~~~~~~~~~~3.0,~~~~~~~~3.5.
\nonumber\\
\label{prodx}
\ea
where we have used the beam  luminosity of $100$ fb$^{-1}$. These  cross
sections  are nearly 10 times larger
than the heavy-quark  pair  production cross section. 
It is clear that even though the cross section decreases rapidly with
increasing fermion mass, 
the number of pairs produced are $ 9\times 10^4(1000)$ even for $m_{F_C}\simeq 1(2)$ TeV.

Even though the pair production cross section
of non-SUSY color-octet fermions is large, unlike gluinos\cite{beenakker}, their decays would be
suppressed in the present minimal Model I. This is because $({45}_F, 1)$ does not have
renormalizable Yukawa type interaction with fermions or Higgs
doublets of $G_{213}\times S_4$ model. Also there are no analogue of
superpartners in this non-SUSY model.      
 Having its origin in the adjoint representation, being neutral under 
$SU(2)_L\times U(1)_Y$,  and in the absence of Yukawa interaction,
the color-octet fermion interacts with quarks at the tree level only
via gluon exchange by which it can hadronize.  
However, the color-octet fermions  may decay
into standard model fermions and one of the light members of $({45}_F,1)$ via higher-dimensional-operator-mediated 
effective interactions
whose strength depends upon possible presence of  scalars with high masses giving rise to 
the operator in some minimally extended models. 
With appropriately longer lifetime, the produced color-octet fermions may
decay outside the detector or with displaced vertices, or some of them may be
even stopped in the detector. Even if no
superpartners are present, they may also form some states, 
analogous to R-hadrons\cite{arkani,baer,farrar2}. These possibilities
 would be explored separately and an extended model may mimic split-
 SUSY model \cite{arkani} to some extent with more interesting collider
 signatures driven by  color-octet fermions. It is clear that at the
 highest LHC energy and a $100$ fb$^{-1}$  luminosity, detection of
 color-octet
pair production would be possible at least up to the particle mass  $\sim 3500$ GeV.    

One of the major goals of LHC and Tevatron is to resolve the issue of supersymmetry
through collider signatures of superpartners  and definite answers
in this respect are expected in the next few years.
In the context of the present flavor symmetric grand unified Model I even without
inclusion of additional GUT representations, collider signature of the 
color-octet fermion would be large production cross section and absence of superpartners.\\  

Two important issues related to the $\sim$TeV mass color-octet fermions are their lifetime and 
relic abundance. Both these are dependent upon the mass $M_{\rm med.}$ of the  scalar
mediators contained in ${\overline {16}_H}\subset SO(10)$ 
generating the effective four-fermion interactions via matter-parity
conserving Yukawa interaction, $Y{45}_F{16}_F{\overline
  {16}_H}$ where $S_4$ quantum numbers have been suppressed and our models
have now been minimally extended to to include the Higgs representation $({\overline
  {16}_H},3)\subset SO(10)\times S_4$. An approximate
formula for the color-octet fermion lifetime is,

\be
 \tau_{F_C}=3\times 10^{-2}sec\left(\frac{M_{\rm med.}}{10^9{\rm GeV}}
\right)^4\left(\frac{1{\rm TeV}}{m_{F_C}}\right)^5
\ee

Normally $M_{\rm med.}$ is expected to be near the GUT-scale or few-orders 
lighter as in Model I, although in Model II it can be even lighter. 
However, we note that 
the $SU(5)$-complete multiplets like ${\overline{10}_H}$ or ${5}_H$ contained
in  ${\overline {16}_H}\subset SO(10)$  
can be made light with any value of $M_{\rm med.}\simeq 10^4$ GeV -$10^{16}$ GeV without 
affecting  coupling unification and 
the GUT-scale ($\sim 10^{16}$ GeV) already achieved in Model I. For example
  cosmologically safe short-lived
color octets  with life-times  $10^{-20}$secs($3\times
10^{-2}$secs) can be easily obtained in the minimal extension of the single-step
breaking model with $M_{\rm med.}=10^4$GeV ($10^9$ GeV) which will be discussed elsewhere while examining collider
signatures. Although introduction of these scalar mediators do not affect the
value of the GUT-scale in the single-step breaking model, they would tend to
increase the value of the GUT-gauge coupling by a small amount with
a correspondingly small  decrease
in the predicted proton-lifetime  by a factor $\sim (35.3\alpha^{\prime}_G)^2$ where 
$\alpha^{\prime}_G$ is the GUT-fine structure constant including the lighter
scalar mediators as would be applicable.  

For a very long-lived octet fermions with lifetime comparable to the age of
the universe, or even larger, corresponding to scalar mediator
masses in the range $10^{13}$ GeV - $10^{16}$ GeV, extensive investigations have been made to
overcome their relic density problem. Perturbatively generated larger relic
density of these color-octet fermions, which may contribute to 
hitherto unobserved DM relic abundance
\cite{arkani,mahasmu}, is usually evaded by invoking second inflation at lower
scale \cite{sinfl}.   
A second possibility  is the substantial reduction of the 
octet-fermion relic density
by rapid pair-annihilations that continue to temperatures much lower than the
freeze-out through various
nonperturbative mechanisms accompanied by Sommerfeld enhancements of the
annihilation cross sections. The general conclusions are that until and unless
the second inflation hypotheses are ruled out or the nonperturbative mechanisms
are proved untenable, long lived octet fermions in the mass range $1-10$ TeV can be
treated cosmologically safe and harmless \cite{baer}.\\   

   Although Tevatron has reached the lower limit 
 $m_{\tilde g} \ge 370 $ GeV  for stopped gluinos,
\cite{khacha}, where interaction with squarks plays 
significant roles, no such limit is available for 
non-supersymmetric color-octet fermions. Similarly, for the conventional gluinos of the
constrained MSSM, CMS collaboration has set the lower bound on gluino mass
$m_{\tilde g} \ge 650$ GeV \cite{cmslt} while ATLAS collaboration has set the lower bound of
 $870$ GeV \cite{ATLAS}, but no such high mass limit is yet available for the 
color-octet fermion of the non-SUSY models.
Very recently a conservative lower bound on the search limit of color- 
octet fermion mass at $m_{F_C}\ge 50$ GeV has been set for LHC energy $\sqrt s
   = 7$ TeV \cite{elberg}. Interesting suggestions have been advanced for
detection of long-lived stopped gluinos with displaced vertices many of which
 are applicable to the case of color-octet fermions discussed in this work \cite{farrar2}.

Pair production of color-octet scalars either through
$q{\bar q}$ annihilation or through gluon-gluon fusion at Tevatron have
been discussed leading to the production cross section of nearly $100$ fb and 
$7$ fb for the scalar masses $250$ GeV and $350$ GeV , respectively. Even
though the production cross section of the color-octet scalars indicated in
the present models are similar, being placed in $({45}_H,2)$, or $({54}_H, 2)$,
or $({210}_H, 2)$ under $SO(10)\times S_4$,  
, their interactions  are somewhat different from those
 discussed in the literature\cite{manohar,grsoctet,soctet}.

\section{Unification with intermediate  symmetry}

In the absence of flavor symmetry, whereas the single-step breaking minimal grand desert models have been  ruled out, GUTs such as $SO(10)$ with one or 
more intermediate symmetries have been  found to be consistent with 
$\sin^2\theta_W$ and proton-lifetime constraints \cite{lmpr,dpar,despande,parpat}. 
In addition, the left-right symmetry breaking intermediate scale has been 
identified with type-I see-saw scale in a number of models \cite{lmpr,parpat}. In 
this section
using flavor unification through $SO(10)\times G_f$ at first
we explore the possibility of  intermediate  Pati-Salam gauge theory with 
unbroken 
D-Parity \cite{dpar,parpat}  and then summarize briefly the 
outcome of other intermediate symmetries.

\subsection{ Intermediate symmetry ${\bf SU(2)_L\times SU(2)_R\times SU(4)_C
(g_{2L}=g_{2R})\times S_4 }$}

\begin{figure}[htb]
\begin{center}
\includegraphics[width=12cm]{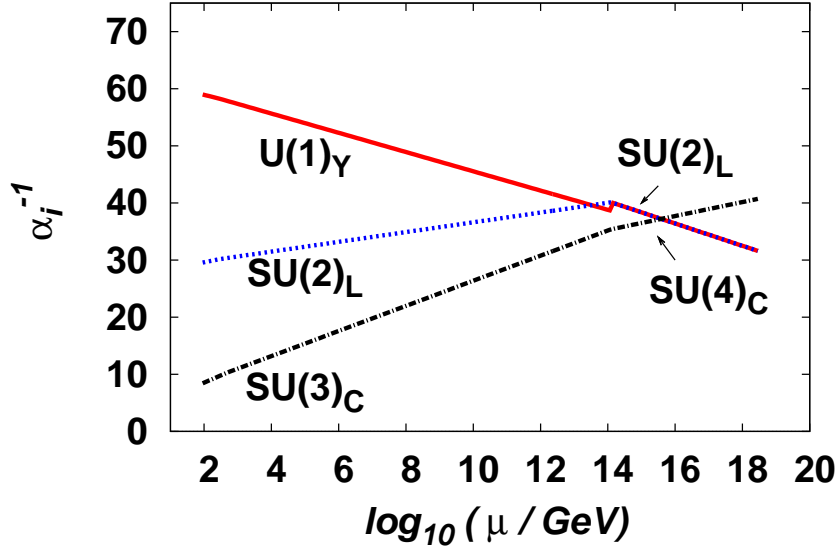}
\caption{ Unification of gauge couplings with $G_{224D}\times S_4$
intermediate symmetry 
shown for Model II, Case (i)  with a
$S_4$- singlet color-octet scalar $C(1,0,8)$
 at $M_C\simeq 500$ GeV described in the text.The unification patterns in Cases
(ii) and (iii) including triplet fermionic DM and color-octet fermion or
 scalar is similar but with different values intermediate and GUT scales.
 }\label{Fig3}
\end{center}
\end{figure}

We consider the symmetry breaking chain

\be  
{\bf {SO(10)\times S_4}} \frac{{\bf M_U}}{} 
~~{\bf  G_{224D}\times S_4} \frac{\bf M_R}{}~~
 {\bf {G_{213}\times S_4}}.\\
\label{2step}
\ee
where we have used the notation $G_{224D}$ for
${\bf SU(2)_L\times SU(2)_R\times SU(4)_C(g_{2L}= g_{2R})}$.
This symmetry has the advantage  that, in the presence of D-parity, 
exact results 
on vanishing corrections on $\sin^2\theta_W$ and intermediate scale $(M_R)$
lead to the stability of $M_R$ \cite{parpat} once it is fixed by the lower
scale parameters in spite of  apprehension that uncertainties on
intermediate scale prediction could be large \cite{sher}. Also  unlike other gauge
theories, $G_{224D}$ has only two gauge couplings which eliminates 
uncertainties in unification that would have otherwise arisen because of the
presence of a triangular region around the GUT scale as is common to three gauge
coupling models.
The first step of breaking is driven assigning  GUT-scale VEV
to the $G_{224}$- singlet in $(54_H, 1) \subset SO(10)\times G_f$ and 
the second step is implemented through $({\overline {126}_H}, 1)\oplus
({45}_H,1)$ under $SO(10)\times G_f$ and the rest are as in Model I. 
Successful unification of couplings with proton stability and right value of 
the see-saw scale is possible if this Model II has  any one of the three 
combinations of particles with masses $\sim$ TeV: (i)
 a  color-octet
scalar, (ii) a color-octet scalar and the fermionic triplet DM, (iii) a 
color-octet fermion and the fermionic triplet DM. The mass parameters and
the GUT-coupling in the first case are,

\underline{Model II}
\ba
& &M_{S}=10^{2.5}~~{\rm GeV}, ~~M_{X}=M_{C}(1,~0,~8)=10^{2.7}~~{\rm GeV},  \nonumber \\
& &~M_R^0=M_{C'}(1,~1,~15)=10^{14.15}~~{\rm GeV},~~ M_U^0=10^{15.6}~~{\rm GeV},~ \alpha_G^{-1}=37.2.
\ea
The RG evaluation and unification of couplings are shown in Fig.\ref{Fig3}.
For the cases (ii) and (iii) the pattern of unification is similar but with
different values of unification and intermediate scales: case (ii) $M_R^0 =
10^{14.6}$ GeV, $M_U^0=10^{15}$ GeV, $ \alpha_G^{-1}=37.1$; case (iii)$M_R^0 =
10^{15.35}$ GeV, $M_U^0=10^{16.9}$ GeV, $ \alpha_G^{-1}=34.6$;

Like Model I, $A_R\simeq 2.347$ in Model II in Case (i).
Since the  proton
lifetime predicted is only a little shorter than the experimental lower bound , 
\ba
\tau_p^0 &\simeq& 10^{-0.3}\times (\tau_p)_{expt.}
\label{eq52}
\ea
it can be easily compensated by small threshold effects. 
We have checked that the $M_S-$threshold effects on the unification mass and
 proton lifetime due to six light scalar doublets vanish.
GUT- threshold effect evaluated including all relevant superheavy scalar 
components in\\ 
$({54}_H, 1), ({45}_H,1), ({\overline{126}_H}, 1)$,\\
 and $({10}_H, 3_1+2+1)$ under
$SO(10)\times G_f$ in the non-degenerate case is,

\ba 
\frac{M_U}{M_U^0}&=&10^{\pm 0.2004|\eta|}.\nonumber\\
\ea
Even if the 
superheavy components are $10(\frac{1}{10})$  times heavier (lighter) than
the GUT scale, the threshold effect gives,

\ba
\tau_p&=&10^{34.0^{+1.54}_{-2.14}}~~ yrs.
\ea
where we have used eq.(13)-eq.(14).
Clearly, this prediction is accessible to ongoing and planned
searches for the decay mode  $p \to e^+\pi^0$.   
 Whereas in Model I 
the Majorana fermion Yukawa coupling $f$ has to be fine tuned by $2-3$ orders
to get the desired heavy RH neutrino mass for the type-I see-saw scale, here we require $f\sim 0.1$. 
 The light
color-octet Higgs scalar with mass $M_C(1,~0,~8) \simeq 500$ GeV - few TeV is
also accessible for detection
at LHC , ILC or other accelerator searches with expectations for remarkable
signatures \cite{grsoctet,soctet}.

We find that when a triplet fermionic DM is used along with
 the complex color-octet scalar as in Case (ii) , exact unification scale is obtained for
$M_U \simeq 10^{15}$ GeV. The deficit in proton lifetime
prediction  by a factor $10^{-2.7}$ below the experimental limit
 can be easily compensated by GUT-scale threshold effects and somewhat
larger splitting with superheavy masses with $\sim 20(1/20)$ times heavier(lighter) than the GUT scale.
 However,
 when both the color-octet fermion
 $F_C(1,0,8)$ and the triplet DM are utilized having TeV scale masses 
 as in Case (iii),
the unification scale rises to ($\simeq 10^{16.9}$ GeV) and the 
GUT-scenario accommodates the triplet DM with a much more stable proton. 
Thus, like the single-step breaking case of Model I, with Pati-Salam
intermediate symmetry too, unification is realizable with low-mass fermions
alone apart from other possibilities.
The derivation of light fermion masses from renormalizable Yukawa interaction 
are carried out in a manner similar to Model I with suitable GUT representations.  

\subsection{ Other intermediate symmetries}

 In this section we briefly state our results obtained using other
 intermediate symmetries like $G_{224}\times S_4(g_{2L}\neq g_{2R})$ and 
$SU(2)_L\times SU(2)_R\times U(1)_{B-L}\times SU(3)_C\times S_4(g_{2L}=g_{2R})
 (\equiv G_{2213D})$.

With $G_{224}\times S_4$ intermediate symmetry and a light color-octet scalar
of mass $M_C= 500$ GeV, we have observed excellent
one-loop unification of couplings at $M_U^0=10^{15.7}$ GeV predicting 
$\tau_p \simeq 1.4\times 10^{34}$~yrs. subject to threshold uncertainties.
But the intermediate scale turns out to be smaller by a factor $(300)^{-1}$
than the desired value of the see-saw scale \cite{hagedorn}. There has been a
recent suggestion to accommodate neutrino masses and mixings with such lower
seesaw scale \cite{buccella}
 
With $G_{2213D}\times S_4$ intermediate symmetry, without using any additional low mass  particles beyond the
minimal requirement of six doublets of the HLM type model,  although we achieve excellent
unification  with right value of $M_R$, the unification scale is found to be
nearly  2 orders smaller than the lower bound imposed by the proton decay
constraint. The other alternative for this model may be that, instead of being
embedded in $SO(10)\times S_4$, it could emerge from a high scale trinification 
model like $SU(3)^3 \times S_4$. Other interesting possibilities through 
flavor-symmetric $SO(10)\times G_f$ will be investigated elsewhere.  

\section{Summary and Conclusion}

In the absence of experimental evidences of supersymmetry, in this work we have 
attempted to implement manifest unification of gauge couplings in the HLM
type model with 
$G_{213}\times S_4$ symmetry restoration at $\sim (1-10)$ TeV through the unifying symmetry  
 $SO(10)\times
G_f$ where $G_f=S_4,~SO(3)_f,$ $SU(3)_f$.
Under the experimental lower bound on proton lifetime and the type-I seesaw
scale constraint, we have carried out completion of grand unification 
successfully in two classes of models: Model I with single-step breaking and
Model II with Pati-Salam intermediate symmetry.\\ 

A special distinguishing feature of the non-SUSY $G_{213}\times S_4$ model with six
doublets at low scales is that it predicts the the color-octet fermion and the
triplet fermionic DM as the non-trivial degrees of freedom missing from MSSM
equivalents
sans squarks and sleptons. As a
result  when these fields are switched on at the TeV scale, the  model 
automatically predicts grand unification of gauge couplings with a non-SUSY
GUT  scale
identical to the MSSM GUT scale ($\sim 10^{16}$ GeV).\\

Compared to the
conventional $SO(10)$ prediction  where the required color-octet 
fermion has been 
found to possess a very large mass, $7\times 10^{10}$ GeV, which can not be
accessed by accelerator searches, in the present model of 
flavor-symmetric $SO(10)\times G_f$, the GUT-signals of both types of exotic
fermions are subject to experimental tests at accelerator energies.      
Both these fermions belonging to the adjoint representation are shown to be 
light due to
suitable values of  the renormalizable Yukawa Lagrangian parameters.
Phenomenology 
of light fermions is briefly outlined. 
Proton lifetime predictions are found to be accessible to ongoing or planned
searches for $p\to e^+\pi^0$.
In two-step breaking model  with Pati-Salam intermediate symmetry, several
possibilities are  pointed out also with experimentally accessible 
proton-lifetime predictions. Current phenomenological investigations suggest
the triplet DM mass in the range $2.75$ TeV - $4.5$ TeV with the predicted positron
excess and absence of anti-proton excess in agreement with indirect DM search
experiments especially when the mass is on the higher side. The color-octet fermion pair  
production cross section and event rate are about one order larger than the
heavy quark pair production  case. These characteristics and absence of
superpartners at LHC energies would point towards the existence of color-octet fermions. With color-octet fermion and the fermionic weak triplet DM  masses being
permitted near  $\sim
1-5$ TeV, more interesting collider signatures are expected in the context of 
flavored grand unification which will be 
investigated separately.

In conclusion, we find that flavor symmetric standard gauge theory can be 
successfully embedded in $SO(10)\times G_f$ theory of femion masses and
unification of three forces with  experimentally testable 
grand unification signals for 
observable proton decay and interesting collider signatures like weak-triplet
DM as well as the color-octet fermion. 

\newpage
\noindent {\bf {Acknowledgment}}
\vskip 0.1 in
The authors acknowledge useful discussions with  K. S. Babu. 
M.K.P. thanks Harish-Chandra Research Institute, Allahabad for a visiting position and
Institute of Physics, Bhubaneswar for facilities.

\end{document}